\documentclass[prl,aps,showpacs,amsmath,amssymb]{revtex4}
\usepackage{amsmath,dcolumn,graphicx,natbib}

\begin{document}

\title{Superconducting gap structure of the 115's revisited}

\author{F. Ronning$^1$, J.-X. Zhu$^1$, Tanmoy Das$^1$, M.J. Graf$^1$, R.C. Albers$^1$, H.B. Rhee$^{1,2}$ and W.E. Pickett$^{2}$}
\address{$^1$Los Alamos National Laboratory, Los Alamos, New Mexico 87545, USA\\
                 $^2$Department of Physics, University of California, Davis, CA 95616, USA
}

%\ead{fronning@lanl.gov}

\date{\today}

\begin{abstract}
Density functional theory calculations of the electronic structure of Ce- and Pu-based heavy fermion superconductors in the so-called 115 family are performed. The gap equation is used to consider which superconducting order parameters are most favorable assuming a pairing interaction that is peaked at ($\pi$,$\pi$,$q_z$) - the wavevector for the antiferromagnetic ordering found in close proximity. In addition to the commonly accepted $d_{x^2-y^2}$ order parameter, there is evidence that an extended s-wave order parameter with nodes is also plausible. We discuss whether these results are consistent with current observations and possible measurements that could help distinguish between these scenarios.
\end{abstract}

% PACS, the Physics and Astronomy
                             % Classification Scheme.
%\keywords{Suggested keywords}%Use showkeys class option if keyword
                              %display desired
%Keywords:anisotropy, magnetoresistance, specific heat, heavy fermions, non-Fermi
%liquid behavior, field tuned quantum phase transition,

\maketitle
%Introduction
%Intro:

CeCoIn$_5$ is a heavy fermion superconductor\cite{petrovic2001}, which has been shown to lie in close proximity to an antiferromagnetic quantum critical point \cite{Pagliuso2002, Zapf2001, Pham06}. It’s structure consists of layers of square planar Ce atoms, similar in that respect to the cuprate superconductors. Soon after the discovery of superconductivity in CeCoIn$_5$ many measurements were performed that were consistent with lines of nodes in the superconducting gap, including specific heat, thermal conductivity \cite{Movshovich01}, spin-lattice relaxation rate \cite{Kohori2001}, and penetration depth \cite{OrmenoPRL2002}. Ideally, ARPES measurements could reveal the anisotropic gap structure in $k$-space, but current equipment has neither the energy resolution nor the temperature range to perform such a study. In its absence probes which are directionally weighted averages over the Fermi surface have been used to identify the location of the nodes, including field-angle dependent specific heat and thermal conductivity\cite{IzawaPRL2001, AokiJPCM2004, VoronstovPRL2006, AnPRL2010}, $H_{c2}(\theta)$\cite{Weickert2006} point contact Andreev reflection measurements\cite{Park115PCARS2008}, and vortex lattice structure \cite{BianchiScience2008, HiasaPRL2008} all of which now argue that nodes lie along the $\Gamma$ $\rightarrow$ ($\pi$,$\pi$,$q_z$) direction in the tetragonal Brillouin zone. In addition, a neutron scattering resonance has been observed at ($\pi$,$\pi$,$\pi$) which can be interpreted as consistent with $d_{x^2-y^2}$ symmetry\cite{Stock2008}.  Taken together the experimental evidence appears fairly compelling that CeCoIn$_5$ is a superconductor with a $d_{x^2-y^2}$ gap structure. Measurements on other members in this family are relatively scarce due to the fact that either pressure is required for superconductivity to occur (as in CeRhIn$_5$, CeIn$_3$, Ce$_2$RhIn$_8$, and CePt$_2$In$_7$\cite{Hegger00, Mathur98, Nicklas2002, Bauer127PRB2010}) or there are radioactive considerations (for PuCoGa$_5$, PuRhGa$_5$, and PuCoIn$_5$\cite{Sarrao2002, Wastin2003, BauerPuCoIn5}). However, with the exception of a few measurements \cite{Shakeripour2007} those measurements which have been made on these compounds are identical to the results found for CeCoIn$_5$ \cite{ParkPRL2008, KasaharaIr115PRL2008, LuIr115Unpub, CurroNature2005, OhishiPRB2007, SarraoJPSJ2007}.

On the theoretical side, the BCS theory of superconductivity allows one to calculate the superconducting gap structure at T = 0 K by solving the gap equation:
\begin{equation*}
\Delta(\mathbf k) = - \sum_{\mathbf k'} \Gamma({\mathbf k},{\mathbf k'}) \frac{\Delta({\mathbf k'})}{\sqrt{|\epsilon_{\mathbf k'}|^2 +|\Delta({\mathbf k'})|^2 }} ,
\label{Eq1}
\end{equation*}
where $\Delta(\mathbf k)$ is the superconducting gap function, $\epsilon_{\mathbf k}$ is the electronic dispersion, and $\Gamma({\mathbf k},{\mathbf k'})$ is the pairing interaction. To solve this equation the low energy electronic structure (ie. the Fermi surface, and effective masses) and the pairing interaction $\Gamma({\mathbf k},{\mathbf k'})$ must be known. This would appear to be a formidable if not impossible task for such strongly correlated materials such as the cuprates or the heavy fermion materials which we wish to study\cite{RainerPhysica1982}. It is well known that for strongly correlated materials there are no well controlled approximations for calculating their properties. However, it does appear from several cases that simple density functional theory (DFT) calculations provide a reasonable starting point for computing Fermi surfaces in good agreement with experiment, even if the mass renormalization can be off by orders of magnitude \cite{SettaiJPCM2001, ShishidoJPSJ2002, HallCeCoIn52001}. The computation of the pairing interaction is more problematic. In weak coupling spin fluctuation theory, however, one may approximate $\Gamma({\mathbf k},{\mathbf k'})$ by a  renormalized spin susceptibility $U\chi_{S}({\mathbf q},\omega )U$ where $U$ is the screened Coulomb potential \cite{Berk1966, MiyakePRB1986, ScalapinoPRB1986, MonodPRB1986, MonthouxNature2007}. Whether or not this is rigorously true in the strongly correlated materials of interest, such as the cuprates, pnictides, and heavy fermions, remains a subject of debate, but at least in the case of the cuprates and the iron based superconductors, the observed gap phenomenology can be reasonably described utilizing this assumption \cite{MonthouxPRL1991,HirschfeldReview2011}.

More specifically, from either a strong coupling, weak coupling, or experimental perspective the spin susceptibility of the cuprates is peaked at ($\pi$,$\pi$). Given the LDA determined Fermi surface with the largest density of states near ($\pm\pi$,0) and (0,$\pm\pi$) the order parameter which will naturally maximize the superconducting gap is a $d_{x^2-y^2}$ order parameter. In the Fe-based superconductors such as doped LaFeAsO and BaFe$_2$As$_2$ the spin susceptibility is peaked at ($\pi$,0). In this case, the combination of a hole pocket centered at the $\Gamma$ point, and an electron pocket at the X point (of the unfolded Brillouin zone) leads to the suggestion of the so-called $s\pm$ gap structure \cite{MazinPRL2008, ChubukovPRB2008, SeoPRL2008, CvetkovicEPL2009}. Furthermore, weak coupling approaches can calculate the variations in the spin susceptibility for different dopings and crystal structures. Inputting the renormalized spin susceptibility and electronic structure leads to strong gap variations and often accidental nodes which are consistent with many of the experimental observations\cite{HirschfeldReview2011}

Since most of the experimental evidence for $d$-wave symmetry in the 115 materials is indirect, the story of the Fe-based superconductors compels us to reconsider the possible superconducting order parameters in the 115 family of superconductors, which are also compensated multiband metals like the Fe-based materials. The goal of this work is not to conclusively show which order parameter is correct in the various 115 materials. Rather, we assume a BCS-like formalism for the superconductivity with a repulsive interaction that is peaked in the vicinity of the nearby long range magnetic order and explore whether the $d$-wave order parameters are the only likely candidates for superconductivity. We find that alternative scenarios are at least plausible particularly for the Pu-based materials. For completeness it must be stated that it is also not obvious whether a BCS picture such as this is even the correct starting point to understand superconductivity in these materials, or whether other ${\mathbf q}$ vectors should be considered for the pairing interaction. However, in the absence of a microscopic theory the phenomenology at least is consistent with what is known about cuprate and Fe-based superconductors, and if it can be shown to be consistent in the heavy fermion superconductors as well, it provides a methodology by which one can tailor higher $T_c$'s within a family of superconductors without detailed knowledge of the pairing mechanism.

\begin{figure}[htbp]
     \centering
\includegraphics[width=3.3in]{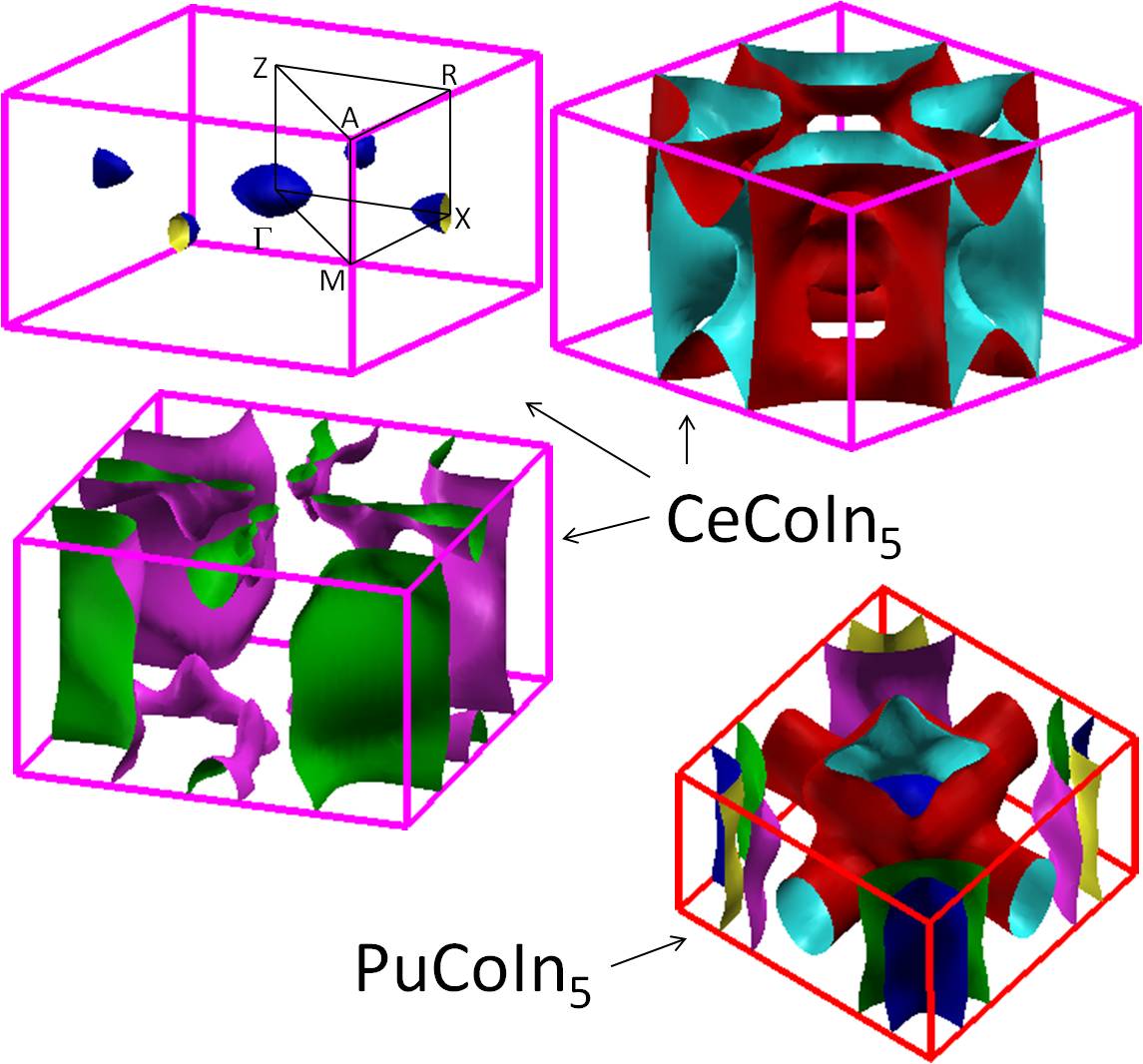}
\caption{\label{CeCoIn5FS}(color online) (top) Three Fermi surface sheets of CeCoIn$_5$. (bottom right) The Fermi surfaces of PuCoIn$_5$.}
\end{figure}

The WIEN2K code\cite{WIEN2K} is used to calculate the electronic structure using the generalized gradient approximation and adopting the Perdew-Burke-Ernzerhof exchange correlation potentials\cite{PBE}. We included spin orbit interactions on the $f$-electrons with a second variational method.

In the Ce-based 115’s the evidence is fairly clear that the spin susceptibility is peaked at ($\pi$,$\pi$,$q_z$). For CeIn$_3$, CeRhIn$_5$, CeCoIn$_5$, Ce$_2$RhIn$_8$, and CePt$_2$In$_7$, it has been shown that a dome of superconductivity emerges around a quantum critical point associated with the suppression of an antiferromagnetically ordered phase with $Q_{AF}$ = ($\pi$,$\pi$,$\delta$)\cite{LawrenceCeIn31980, Benoit1980, Bao2000, Nicklas2007, Bao218PRB2001, apRobertsPRB2010, Sakai127PRB2011}. We expect a similar situation to exist for Ce$_2$PdIn$_8$ and Ce$_2$CoIn$_8$\cite{Ce2PdIn8, Ce2CoIn8}.  Furthermore, the neutron resonance observed in the superconducting state of CeCoIn$_5$ \cite{Stock2008}, is either a result of the gapping of the spin excitations present at $Q$ = ($\pi$,$\pi$,$\pi$)\cite{Chubukov115PRL2008} or a consequence of the pairing interaction being peaked at this wavevector\cite{EreminPRL2008}. Consequently, assuming a uniform $U$ the pairing interaction is indeed peaked at $Q$ = ($\pi$,$\pi$,$q_z$) for all of the Ce-based superconductors listed above.

In Fig. \ref{CeCoIn5FS}, we show the Fermi surface for CeCoIn$_5$. It is a very complicated 3D Fermi surface. Tight binding models often approximate the Fermi surface in the 115’s by multiple cylindrical electron Fermi surfaces centered at ($\pi$,$\pi$,$k_z$) and a hole like Fermi surface centered at $\Gamma$(e.g. \cite{MaehiraJPSJ2003}). For the purpose of discussion a cartoon of the resulting Fermi surface is shown in Fig. 2 for $k_z$=0.

\begin{figure}[htbp]
     \centering
     \includegraphics[width=0.5\textwidth]{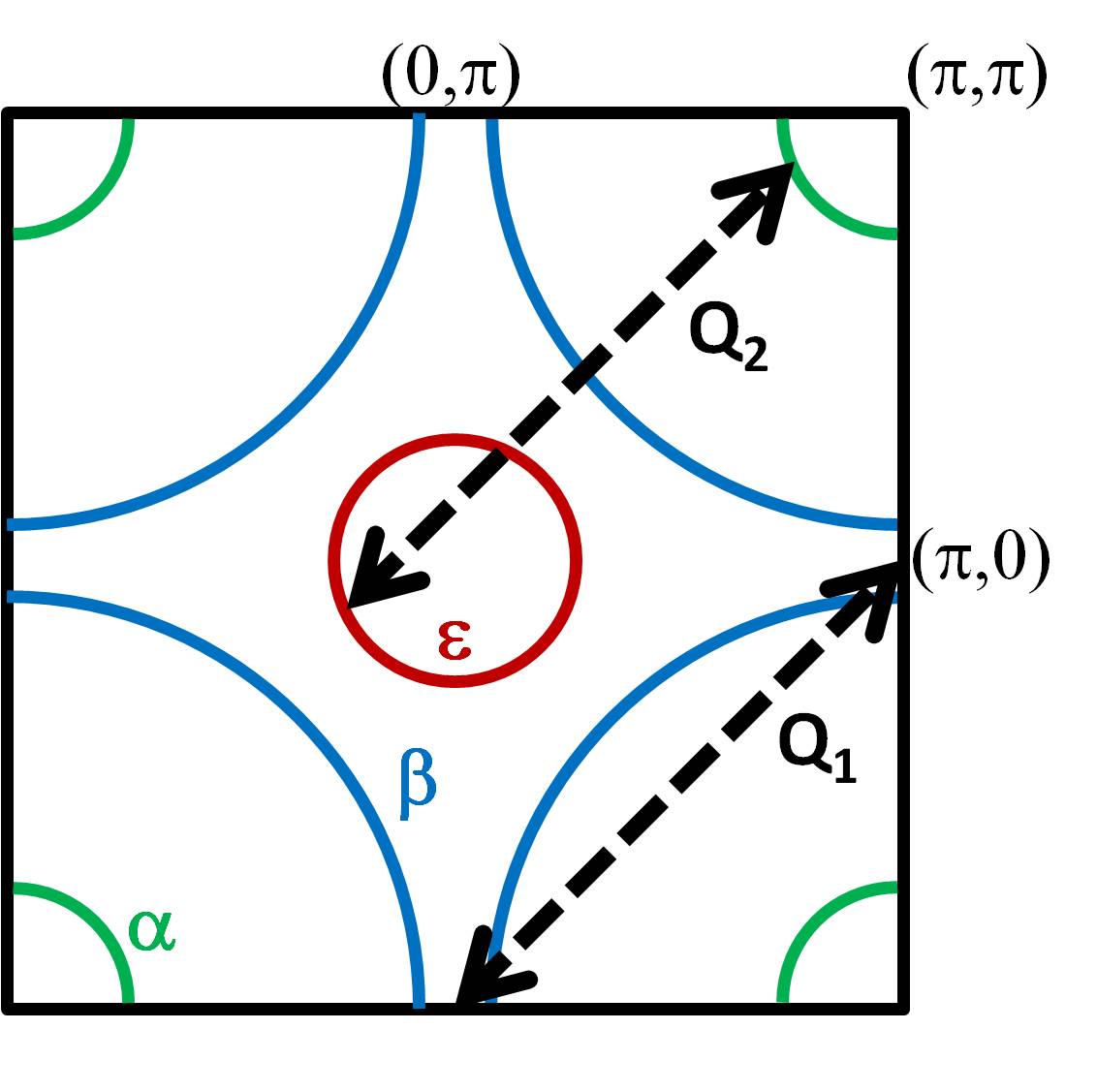}
     \caption{(color online) Cartoon Fermi surface of the 115's in the ($k_x$, $k_y$) plane. The wavevector $Q$ = ($\pi$,$\pi$) represents where the pairing fluctuations are likely maximal assuming a constant $U$ and orbital matrix elements. Depending on which states they dominantly pair (ie. $Q_1$ or $Q_2$) will dictate which order parameter symmetry is most favorable.}
\label{FSCartoon}
\end{figure}

We now ask the question of what gap symmetry is favored by such a Fermi surface with a pairing potential that is peaked at ($\pi$,$\pi$,$q_z$). From Eq. (1) the order parameter will be chosen so that the maximal number of states within the gap energy of the Fermi energy which are separated by the wave vector $Q$ = ($\pi$,$\pi$,$q_z$) can have an order parameter with opposite sign (ie. $\Delta(\mathbf k)$ = - $\Delta(\mathbf k + Q)$). As in the cuprates, this results in a $d_{x^2-y^2}$ symmetry when a large DOS exists at ($\pi$,0,$k_z$) and equivalent points. This certainly is a possibility for the 115’s as emphasized in Fig. 2 by the vector $Q_1$. However, a second possibility is that the fluctuations strongly pair the states on different Fermi surface sheets (e.g. those sheets labeled $\alpha$ and $\epsilon$ in Fig. 2 with vector $Q_2$). This situation is analogous to the Fe-based superconductors\cite{MazinPRL2008, ChubukovPRB2008, SeoPRL2008, CvetkovicEPL2009}. Consequently, an $s\pm$ type pairing symmetry, belonging to the $A_{1g}$ representation where the order parameter changes sign between different Fermi surface sheets, would be favored. Given the complexity of the actual Fermi surface, and the need to overcome the strong onsite Coulomb repulsion $U$ \cite{ChubukovPRB2009, MaierPRB2009} it is likely that accidental nodes exist if the $s\pm$ order parameter was the dominant order parameter. If the coupling between Fermi surface sheets is sufficiently weak then it is also possible to have different symmetry representations on different Fermi surface sheets (ie. $s\pm$ on $\alpha$ and $\gamma$ and $d_{x^2-y^2}$ on $\beta$) \cite{RourkePRL2005,SeyfarthPRL2008}.

As an initial investigation to determine whether an $s\pm$ or a $d_{x^2-y^2}$ order parameter is more likely to occur in any particular 115 compound we examine the band structure more closely. One should consider the states which once renormalized to account for the enhanced Sommerfeld coefficient of the specific heat are within the spin fluctuation energy of $E_F$. dHvA measurements show that the experimentally measured Fermi surfaces are in good agreement with those computed using non-interacting DFT theory \cite{SettaiJPCM2001}. Thus, even though the mass renormalization is not correct, the DFT computed electronic structure should give a reasonable first guess as to the leading pairing instability. We suggest that the $d_{x^2-y^2}$ order parameter will be dominant when the majority of states are found near ($\pi$,0,$k_z$) and equivalent points, while the $s\pm$ order parameter will be favored when the majority of states are found near the (0,0,$k_z$) and ($\pi$,$\pi$,$k_z$) points. For completeness all representations consistent with the tetragonal symmetry of the lattice should be considered\cite{SigristRMP1991}, but given the crudeness of this analysis for simplicity we only consider the $d_{x^2-y^2}$ and $s\pm$ order parameters as the most likely scenarios.

\begin{figure}[htbp]
     \centering
     \includegraphics[width=0.49\textwidth]{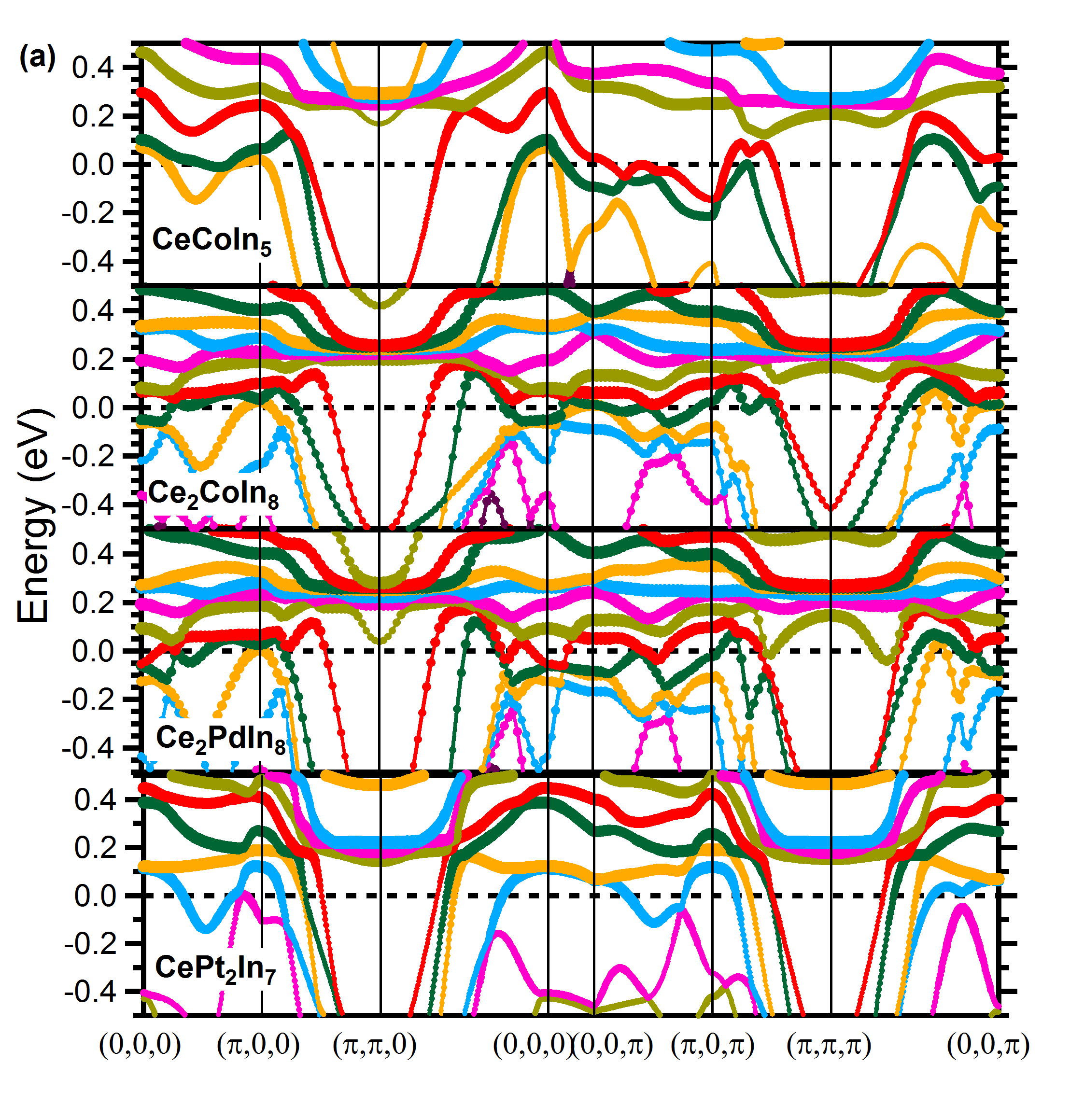}
     \includegraphics[width=0.49\textwidth]{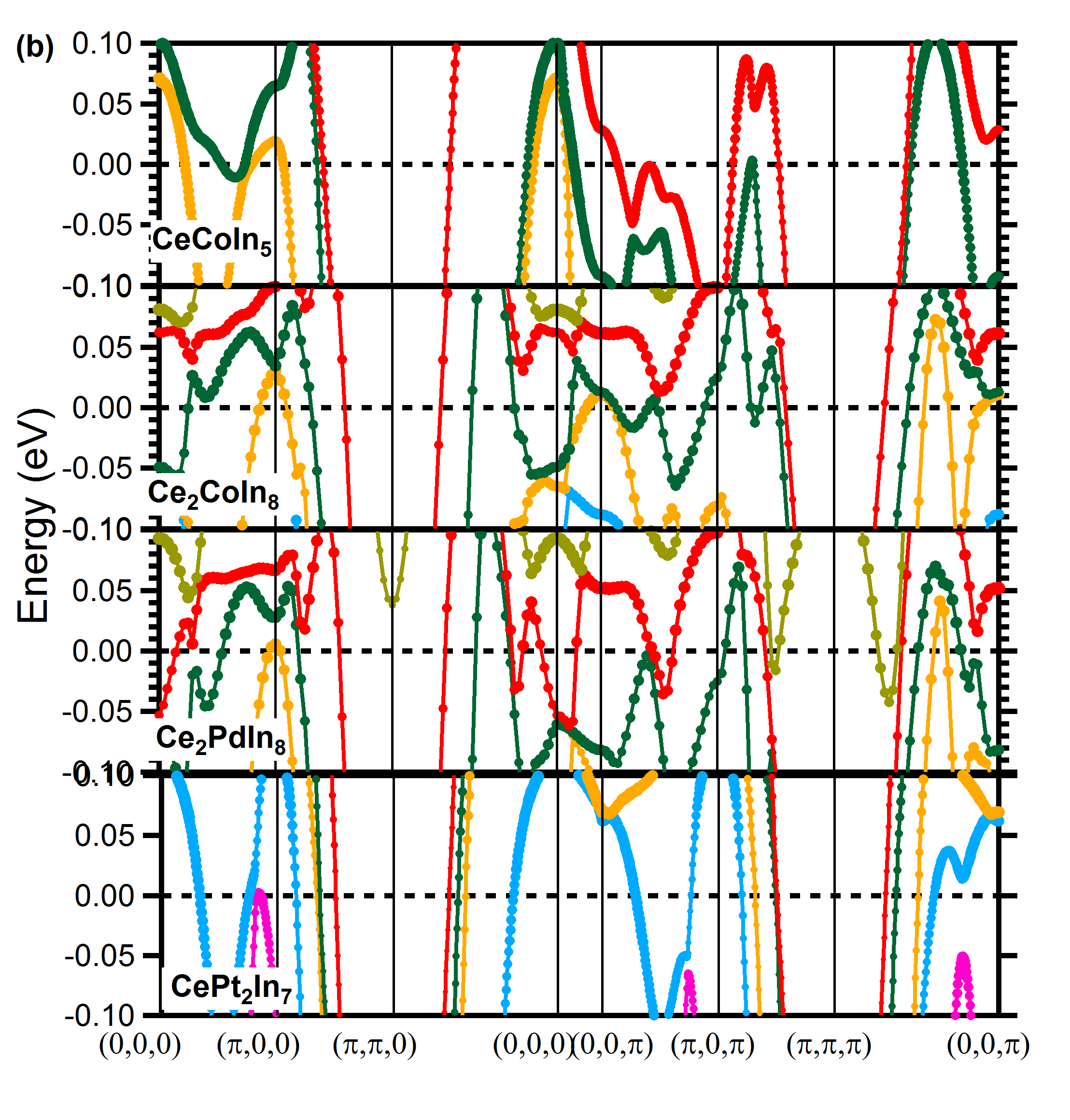}
\caption{\label{CeBS}(color online) (a) Band structure from top to bottom of CeCoIn$_5$, Ce$_2$CoIn$_8$, Ce$_2$PdIn$_8$ and CePt$_2$In$_7$. The size of the points reflects the amount of $f$-character. (b) Zoom of panel (a). }
\end{figure}

Fig. 3 shows the band structure for several Ce-based superconducting materials, which are consistent with other band structure calculations where available \cite{MaehiraJPSJ2003, SettaiJPCM2001}. For the 115’s there are states near $\Gamma$ and a large cylindrical Fermi surface centered at ($\pi$,$\pi$,$k_z$) which could favor an $s\pm$ order parameter, but especially when considering the entire $k_z$ dispersion, there are significantly more states available for superconductivity near ($\pi$,0,$k_z$) and equivalent points which would indeed favor $d_{x^2-y^2}$ pairing. A non-uniform renormalization in $k$-space or strong variations in orbital character could easily shift the preference to an alternative order parameter. dHvA results  indeed indicate that the mass renormalization is nonuniform across momentum space. However, they show that the small pocket around the $\Gamma$ point is actually the lightest band \cite{SettaiJPCM2001} which thus further supports the $d_{x^2-y^2}$ gap structure over $s\pm$. While theoretical studies of the order parameter in the 115's generally favor $d_{x^2-y^2}$ symmetry \cite{TanakaJPSJ2006, FukuzawaJPSJ2003}, a study of the orbitally degenerate Hubbard model with tight binding parameters for the 115s also found alternative representations depending on the Coulomb and crystal field parameters \cite{TakimotoJPCM2002}.

We also show the band structures for Ce$_2$PdIn$_8$, Ce$_2$CoIn$_8$, and CePt$_2$In$_7$ in Fig. 3.  The 218's can be visualized as a bilayer variant of the 115 structure. Indeed near 0.2 eV a doubling of the number of bands is clearly visible (especially clear near ($\pi$,$\pi$,$k_z$)). However, the number of Fermi surface sheets remains unchanged and the electronic structure near $E_F$ is qualitatively similar between  Ce$_2$CoIn$_8$ and  CeCoIn$_5$. The electronic structure of Ce$_2$PdIn$_8$ near $E_F$ is similar to Ce$_2$CoIn$_8$ with a small shift of the chemical potential reflecting the extra electron of Pd relative to Co, although it is clearly not a rigid shift. Finally, for CePt$_2$In$_7$, the additional PtIn$_2$ bands relative to the 115 structure does result in additional Fermi surface sheets, although quite similar in structure to the other large Fermi surface sheets\cite{AltarawnehCePt2In7PRB2011}. While the details vary, the main features observed in the  band structure of CeCoIn$_5$ persist in these three other compounds as well. Namely, the dominant weight is found close to the ($\pi$,0,$k_z$) points suggesting dominantly $d_{x^2-y^2}$ gap symmetry.  The magnetism and superconductivity present in the 115's appears robust to these minor changes in the electronic structure.

In Fig. 4 we show the analogous band structures for the Pu-based superconductors, again in agreement with previous publications \cite{OpahlePRL2003, MaehiraPRL2003, ZhuPuCoIn5, ElgazzarPu218s}. As seen in Fig. 1, the Fermi surface of PuCoIn$_5$ is qualitatively similar to CeCoIn$_5$, with respect to cylindrical-like Fermi surface sheets centered at ($\pi$,$\pi$,$q_z$), a small hole pocket near $\Gamma$ and a large Fermi surface sheet. However, one can also tell that in PuCoIn$_5$, the hole pocket is larger, the cylindrical sheets are closer to ($\pi$,$\pi$,$q_z$), and there is less weight near  ($\pi$,0,$q_z$) and equivalent points. Note, that in the case of the Pu-based superconductors it is much less clear where to guess the pairing fluctuations might be peaked since the analogous long range order has yet to be found. However, for the purpose of this discussion we will again assume that the pairing fluctuations are peaked at ($\pi$,$\pi$,$q_z$). This is indeed what one would assume on the basis of a uniform $U$ spin fluctuation calculation \cite{Das}. In contrast to the case of the Ce-based superconductors the arguments for $d_{x^2-y^2}$ symmetry are much less conclusive. Relative to the Ce-based materials there are fewer states in the vicinity of ($\pi$,0,$k_z$) and more states near the Fermi level at ($\pi$,$\pi$,$k_z$). This leads us to speculate that the potential for an $s\pm$ order parameter is greater in the case of Pu-based 115’s than in the case of Ce-based 115’s. Note also that in contrast to the Ce-based compounds, the bare $f$ level is situated directly at the Fermi level in these Pu compounds. As a consequence, the bilayer variants such as Pu$_2$CoGa$_8$ have dramatically different electronic structure relative to PuCoGa$_5$. This is a possible reason for why superconductivity is much less prevalent in the relatives of the Pu-based 115s \cite{ElgazzarPu218s, RheePuPt2In7}, while it is much more robust in the Ce-based 115's.

\begin{figure}[htbp]
     \centering
\includegraphics[width=0.49\textwidth]{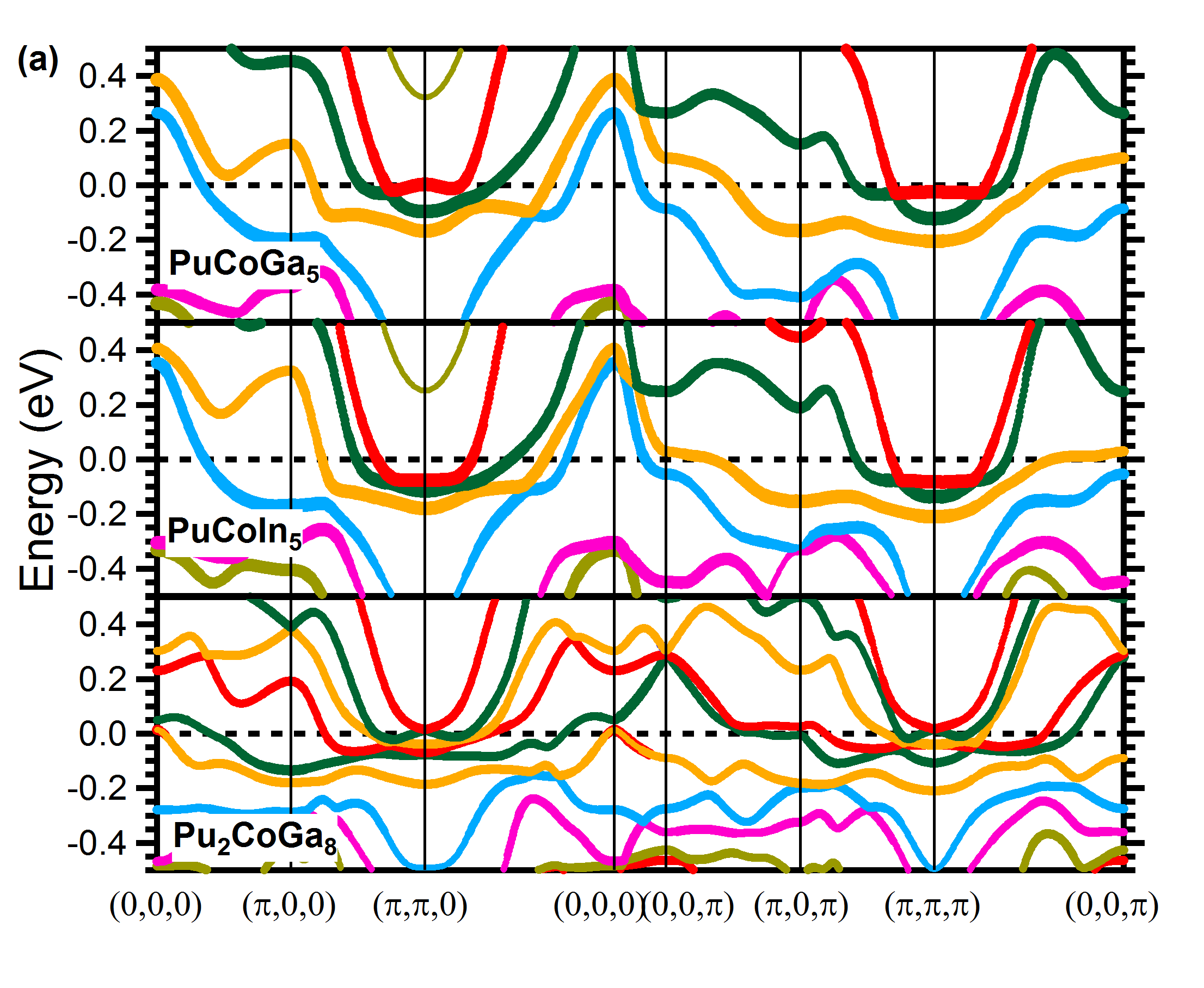}  % [width=3.3in]
\includegraphics[width=0.49\textwidth]{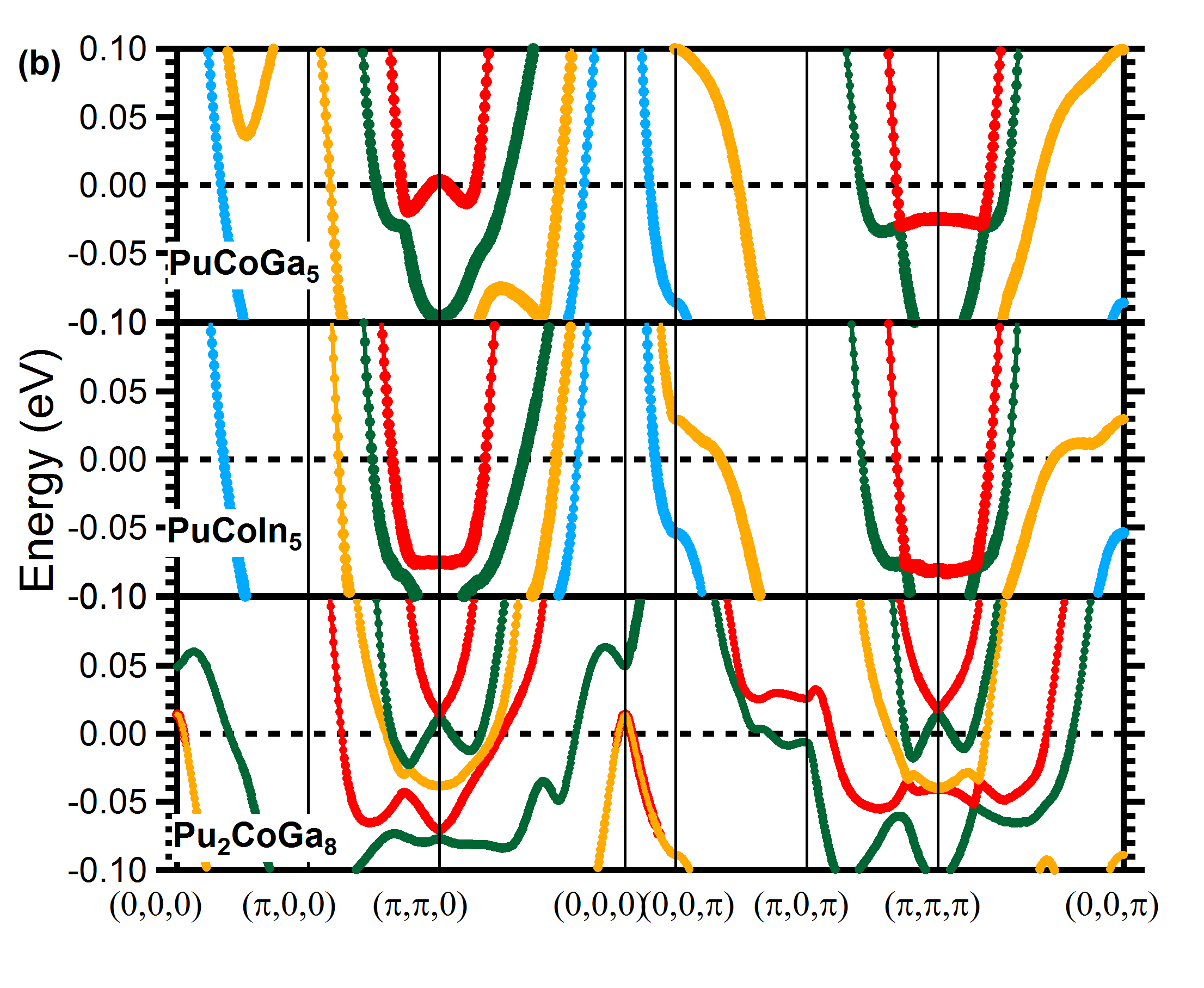}
\caption{\label{PuBS}(color online) (a) Band structure from top to bottom of PuCoGa$_5$, PuCoIn$_5$ and Pu$_2$CoGa$_8$ (which is not superconducting). The size of the points reflects the amount of $f$-character.  (b) Zoom of panel (a) }
\end{figure}

We now briefly reconsider the experimental evidence for the position of the nodes in these materials. Ideally, field angle specific heat and thermal conductivity measurements can give the position of the nodes. However, they rely on the ability to accurately model the true electronic structure, which as we have emphasized, is poorly understood in heavy fermion materials. Furthermore, a similar 4-fold modulation has also been reported in some of the Fe-based superconductors \cite{YamashitaPRB2011, ZengNatComm2010}. Similarly, while the point contact Andreev spectroscopy data could differentiate in favor of a $d_{x^2-y^2}$ order parameter relative to a $d_{xy}$ order parameter, more work is needed for a realistic electronic structure whether alternative gap structures including $s\pm$ with accidental nodes are possible \cite{Park115PCARS2008,FogelstromPRB2010}. We also find it interesting that the universal limit of the residual term in the thermal conductivity \cite{GrafPRB1996} is not obeyed with La-doping in CeCoIn$_5$ \cite{TanatarPRL2005}. This likely reflects the multiband nature of superconductivity in these materials \cite{BarzykinPRB2007}.

What experiments could help resolve the question of the order parameter in these systems? Based on similar work which is being proposed to differentiate various gap symmetries in the Fe-based superconductors it is easy to propose similar measurements for the Ce- and Pu-based 115 superconductors, even though we acknowledge that temperature, pressure, and radioactive constraints may make many of these measurements very difficult. Certainly, phase sensitive experiments as done on cuprates \cite{VanHarlingenRMP1995, TsueiRMP2000} would clearly be the most definitive tests of the order parameter. Additional studies such as examining details of impurity states with scannning tunneling microscopy\cite{JianxinSTMPRL2011} and the quasiparticle interference pattern\cite{QPI115}, the neutron resonance \cite{DasPRL2011}, and Andreev bound states \cite{GhaemiPRL2009} are a few examples of probes that will further help distinguish between different possible order parameters. In the exceptional case of the 20 K PuCoGa$_5$ superconductor, ARPES measurements are obviously also potentially enlightening. Note, that one should also consider the possibility that the order parameter may change (and could gain/lose nodes) as a function of tuning parameters such as pressure or doping \cite{ParkNJP2009}, as has been suggested in the Fe-based superconductors.

%Conclusion:

We have examined the band structures for various Ce- and Pu-based superconductors within the 115 family. Based on the assumption of pairing fluctuations which are peaked at $q_x$ = $q_y$ = $\pi$ we find that the $d_{x^2-y^2}$ order parameter is the likely candidate for superconductivity in the Ce-based materials. However, we also find that the $s\pm$ order parameter may be favored if an anisotropic mass renormalization favors the gapping between states in the vicinity of (0,0,$q_z$) and ($\pi$,$\pi$,$q_z'$). Experimentally this does not appear to be the case in CeCoIn$_5$, but it is a much stronger possibility for the Pu-based superconductors. The purpose of this survey is to stimulate further discussion of the order parameter in the 115 based superconductors. Future theoretical and experimental work is needed to clarify the order parameter in these materials.

\section*{Acknowledgements}

Work at Los Alamos National Laboratory was
performed under the auspices of the U.S. Department of Energy partially under Contract No. DE-AC52-06NA25396, the Office of Science (BES), Los Alamos LDRD, and a NERSC computing allocation under Contract No. DE-AC02-05CH11231. H. B. R. and W. E. P. were supported by DOE SciDAC-e grant
DE-FC02-06ER25777.

\section*{References}
%\bibliography{RonningSCES2011_bib}

%merlin.mbs apsrev4-1.bst 2010-07-25 4.21a (PWD, AO, DPC) hacked
%Control: key (0)
%Control: author (8) initials jnrlst
%Control: editor formatted (1) identically to author
%Control: production of article title (-1) disabled
%Control: page (0) single
%Control: year (1) truncated
%Control: production of eprint (0) enabled
%

\end{document}